\documentclass[twocolumn,10pt]{IEEEtran}

\usepackage{amsmath,amssymb,amsthm}
\usepackage{graphicx}
\usepackage{xcolor}
\usepackage[acronym]{glossaries}
\usepackage{url}

\newacronym{ris}{RIS}{reconfigurable intelligent surface}
\newacronym{d-ris}{D-RIS}{diagonal RIS}
\newacronym{bd-ris}{BD-RIS}{beyond-diagonal RIS}
\newacronym{siso}{SISO}{single-input single-output}
\newacronym{simo}{SIMO}{single-input multiple-output}
\newacronym{miso}{MISO}{multiple-input single-output}
\newacronym{mimo}{MIMO}{multiple-input multiple-output}
\newacronym{los}{LoS}{line-of-sight}
\newacronym{snr}{SNR}{signal-to-noise ratio}
\newacronym{svd}{SVD}{singular value decomposition}
\newacronym{iid}{i.i.d.}{independent and identically distributed}

\newtheorem{definition}{Definition}
\newtheorem{lemma}{Lemma}
\newtheorem{theorem}{Theorem}
\newtheorem{corollary}{Corollary}

\begin{document}

\title{\huge Can Reciprocal RIS Perform as Non-Reciprocal RIS?}

\author{Matteo~Nerini,~\IEEEmembership{Senior~Member,~IEEE},
        Juan Vidal Alegría,~\IEEEmembership{Member,~IEEE},
        Bruno~Clerckx,~\IEEEmembership{Fellow,~IEEE}

\thanks{This work has been supported in part by UKRI under Grant EP/Y004086/1, EP/X040569/1, EP/Y037197/1, EP/X04047X/1, EP/Y037243/1.}
\thanks{Matteo Nerini and Bruno Clerckx are with the Department of Electrical and Electronic Engineering, Imperial College London, SW7 2AZ London, U.K. (e-mail: m.nerini20@imperial.ac.uk; b.clerckx@imperial.ac.uk).}
\thanks{Juan Vidal Alegría is with the Department of Electrical and Information Technology, Lund University, 221 00 Lund, Sweden. (e-mail: juan.vidal\_alegria@eit.lth.se).}}

\maketitle

\begin{abstract}
Yes, if and only if some optimal non-reciprocal RIS (i.e., maximizing a given performance metric) makes the round-trip channel transmitter-RIS-transmitter symmetric, which holds if the channel transmitter-RIS has rank one or if single-stream transmission is optimal.
\end{abstract}

\begin{IEEEkeywords}
Non-reciprocal RIS, reciprocal RIS, reconfigurable intelligent surface (RIS).
\end{IEEEkeywords}

\section{Introduction}

\begin{figure*}[t]
\centering
\includegraphics[width=0.98\textwidth]{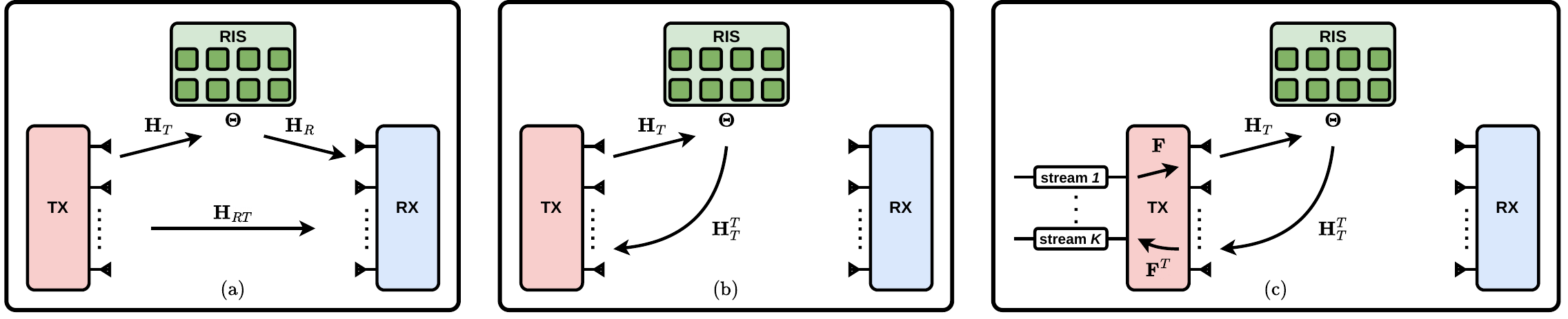}
\caption{RIS-aided system with $N_T$ transmitting antennas and $N_R$ receiving antennas where we highlight:
(a) the $N_R\times N_T$ channel from the $N_T$ transmitting to the $N_R$ receiving antennas $\mathbf{H}_{RT}+\mathbf{H}_R\mathbf{\Theta}\mathbf{H}_T$;
(b) the $N_T\times N_T$ round-trip channel from the $N_T$ transmitting antennas to themselves $\mathbf{H}_T^T\mathbf{\Theta}\mathbf{H}_T$;
(c) the $K\times K$ round-trip channel from the $K$ transmitted streams to themselves $\mathbf{F}^T\mathbf{H}_T^T\mathbf{\Theta}\mathbf{H}_T\mathbf{F}$.}
\label{fig:system}
\end{figure*}

\Gls{ris} has been largely explored as a technology that can enhance wireless communications by shaping the wireless channel \cite{wu21,zha20}.
Consider a generic wireless system between $N_T$ transmitting antennas and $N_R$ receiving antennas, as shown in Fig.~\ref{fig:system}(a).
The wireless channel $\mathbf{H}\in\mathbb{C}^{N_R\times N_T}$ with an $M$-element \gls{ris} is given by
\begin{equation}
\mathbf{H}=\mathbf{H}_{RT}+\mathbf{H}_R\mathbf{\Theta}\mathbf{H}_T,\label{eq:H}
\end{equation}
where $\mathbf{H}_{RT}\in\mathbb{C}^{N_R\times N_T}$ is the direct channel between the transmitting and receiving antennas, $\mathbf{H}_R\in\mathbb{C}^{N_R\times M}$ is the channel between the \gls{ris} and the receiving antennas, $\mathbf{\Theta}\in\mathbb{C}^{M\times M}$ is the \gls{ris} reflection matrix, and $\mathbf{H}_T\in\mathbb{C}^{M\times N_T}$ is the channel between the transmitting antennas and the \gls{ris} \cite{wu21}.
In this system, the $N_T$ transmitting antennas can belong to a single or multiple transmitters, and, similarly, the receiving antennas can belong to a single or multiple receivers.
We can control the wireless channel $\mathbf{H}$ with a \gls{ris} by optimizing its reflection matrix $\mathbf{\Theta}$, which formally is the scattering matrix of the microwave network implementing the \gls{ris}.
The reader is referred to \cite[Chapter~4]{poz11} for more details on the physical meaning of a scattering matrix.

Conventionally, \gls{ris} has been designed by controlling each element independently through a tunable load \cite{wu21,zha20}.
Assuming lossless loads, this design results in a scattering matrix
\begin{equation}
\mathbf{\Theta}=\text{diag}(e^{j\theta_1},e^{j\theta_2},\ldots,e^{j\theta_M}),\label{eq:T-D}
\end{equation}
where $\theta_m\in[0,2\pi)$, for $m=1,\ldots,M$.
Given the structure of this $\mathbf{\Theta}$, such \gls{ris} design has been denoted as \gls{d-ris} \cite{li26}.
Nevertheless, more general \gls{ris} designs are possible, which have a scattering matrix $\mathbf{\Theta}$ not limited to being diagonal and go under the umbrella term of \gls{bd-ris} \cite{li26,she22}.
By implementing a \gls{ris} with an arbitrary lossless and reciprocal microwave network, it is possible to optimize $\mathbf{\Theta}$ subject to unitary (due to losslessness) and symmetric (due to reciprocity) constraints, i.e.,
\begin{equation}
\mathbf{\Theta}^H\mathbf{\Theta}=\mathbf{I}_M,\;\mathbf{\Theta}=\mathbf{\Theta}^T,\label{eq:T-R}
\end{equation}
which are the most general constraints of a lossless \textit{reciprocal \gls{ris}} \cite{li26,she22,ale25}.
Note that the constraints of a reciprocal \gls{ris} in \eqref{eq:T-R} are more general than the constraints of a \gls{d-ris} in \eqref{eq:T-D}, and therefore a reciprocal \gls{ris} performs at least as well as a \gls{d-ris}.
Even more general \gls{ris} designs have been proposed, including the use of non-reciprocal microwave components such as circulators, isolators, and gyrators \cite{ale25,bar23,xu25}.
In particular, by implementing a \gls{ris} with an arbitrary lossless microwave network, $\mathbf{\Theta}$ is only constrained to be unitary, i.e.,
\begin{equation}
\mathbf{\Theta}^H\mathbf{\Theta}=\mathbf{I}_M,\label{eq:T-NR}
\end{equation}
giving the most general constraints of a lossless \textit{non-reciprocal \gls{ris}}.
The terminology non-reciprocal \gls{ris} does not mean that the \gls{ris} is not reciprocal.
It means that the \gls{ris} is not required to be reciprocal, i.e., it is not constrained to have a symmetric $\mathbf{\Theta}$, consistent with previous literature \cite{ale25,bar23,xu25}.
Since the constraints of a non-reciprocal \gls{ris} in \eqref{eq:T-NR} are more general than those of a reciprocal \gls{ris} in \eqref{eq:T-R}, a non-reciprocal \gls{ris} performs at least as well as a reciprocal \gls{ris}.

Related literature explored non-reciprocal \gls{ris} mainly for uplink-downlink channel reciprocity attacks \cite{xu25,wan24,wan26,wan25} and for full-duplex systems \cite{li25,liu26-a,liu26-b,liu26-c}.
In addition, \cite{ale25} showed that a reciprocal \gls{ris} requires more elements than a non-reciprocal one to arbitrarily reconfigure the channel, e.g., for channel orthogonalization.
However, non-reciprocal \gls{ris} requires non-reciprocal components (e.g., circulators, isolators, and gyrators), which increase hardware complexity.
It is therefore essential to understand when this additional complexity brings a performance gain.
In this paper, we therefore answer the fundamental question: \textit{Can reciprocal \gls{ris} perform as non-reciprocal \gls{ris}?}
Our contributions are as follows.

\textit{First}, we derive a necessary and sufficient condition for reciprocal \gls{ris} to perform as non-reciprocal \gls{ris} with the same number of elements.
This condition states that there must exist an optimal non-reciprocal \gls{ris} (i.e., maximizing the considered performance metric) that makes the round-trip channel from the transmitting antennas, via the \gls{ris}, back to the transmitting antennas symmetric.
We then show that, as a consequence of this condition, reciprocal \gls{ris} performs as non-reciprocal \gls{ris} when the channel from the transmitting antennas to the \gls{ris}, or from the \gls{ris} to the receiving antennas, has rank one.
This includes all single-user \gls{siso}, \gls{simo}, and \gls{miso} systems.

\textit{Second}, we refine this necessary and sufficient condition under linear precoding, which is not assumed in our first contribution.
In this case, symmetry is only required for the round-trip channel seen by the transmitted streams.
Since this round-trip channel is a scalar for single-stream transmission, reciprocal \gls{ris} performs as non-reciprocal \gls{ris} when single-stream transmission is optimal.

The statements of all lemmas, theorems, and corollaries have been formally verified in Lean~4 with the assistance of Claude (Anthropic), and the code is available at:
\\\url{https://github.com/matteonerini/reciprocal-ris}.

\section{Reciprocal Versus Non-Reciprocal RIS}

In this section, we formally define when reciprocal \gls{ris} performs as non-reciprocal \gls{ris}, and we derive two lemmas characterizing the operations that each of them can realize.

Let $\mathcal{U}=\{\mathbf{\Theta}\in\mathbb{C}^{M\times M}:\mathbf{\Theta}^H\mathbf{\Theta}=\mathbf{I}_M\}$ and $\mathcal{S}=\{\mathbf{\Theta}\in\mathcal{U}:\mathbf{\Theta}=\mathbf{\Theta}^T\}$ denote the sets of feasible scattering matrices of a non-reciprocal and a reciprocal \gls{ris}, respectively.
To compare these two architectures in any wireless system under any performance metric, we first observe how a \gls{ris} affects performance.
The transmitted signals reach the \gls{ris} only through the channel $\mathbf{H}_T$, and the signals reflected by the \gls{ris} reach the receivers only through the channel $\mathbf{H}_R$.
Therefore, most, if not all, performance metrics depend on $\mathbf{\Theta}$ through $\mathbf{\Theta}\mathbf{H}_T$, through $\mathbf{H}_R\mathbf{\Theta}$, or through both.
In this work, we consider metrics of the form $f(\mathbf{\Theta}\mathbf{H}_T)$, where $f$ can be any real-valued metric, such as capacity, sum-rate, max-min fairness, or energy efficiency, in which all other design variables, such as precoders and combiners, have already been optimized.
For metrics of the form $f(\mathbf{H}_R\mathbf{\Theta})$, all results of this work hold with $\mathbf{H}_T$ replaced by $\mathbf{H}_R^T$, since $\mathbf{H}_R\mathbf{\Theta}=(\mathbf{\Theta}^T\mathbf{H}_R^T)^T$ and both $\mathcal{U}$ and $\mathcal{S}$ are closed under transposition.
Metrics depending on $\mathbf{H}_R\mathbf{\Theta}\mathbf{H}_T$, such as any function of the channel $\mathbf{H}$, are of both forms, so all results of this work apply to them both as they are and with $\mathbf{H}_T$ replaced by $\mathbf{H}_R^T$.
Note that metrics that are functions of $\mathbf{\Theta}\mathbf{H}_T$ but not of the whole $\mathbf{H}_R\mathbf{\Theta}\mathbf{H}_T$ are rare but exist.
For example, in a two-operator system \cite{wu26}, a \gls{ris} controlled by one operator must not alter the channels of the other operator, imposing a fixed $\mathbf{\Theta}\mathbf{h}_{T,2}$ in \cite[Equation~(2)]{wu26}.
These constraints involve $\mathbf{\Theta}\mathbf{H}_T$ but not the whole $\mathbf{H}_R\mathbf{\Theta}\mathbf{H}_T$.

We assume that the maximum of the considered metric over $\mathcal{U}$ is attained, which holds, e.g., when the metric is continuous, since $\mathcal{U}$ is compact.
Since $\mathcal{S}\subset\mathcal{U}$, a non-reciprocal \gls{ris} always performs at least as well as a reciprocal \gls{ris}.
We are interested in when a reciprocal \gls{ris} is enough to achieve optimal performance, as formalized by the following definition.
\begin{definition}
\label{def:perf}
We say that reciprocal \gls{ris} performs as non-reciprocal \gls{ris} when there exists $\mathbf{\Theta}^\star\in\mathcal{S}$ such that
\begin{equation}
f\left(\mathbf{\Theta}^\star\mathbf{H}_T\right)=
\max_{\mathbf{\Theta}\in\mathcal{U}} f\left(\mathbf{\Theta}\mathbf{H}_T\right).\label{eq:perf}
\end{equation}
\end{definition}

Definition~\ref{def:perf} suggests a simple strategy to verify whether reciprocal \gls{ris} can perform as non-reciprocal \gls{ris}.
Let $\mathbf{\Theta}_U^\star\in\mathcal{U}$ be an optimal non-reciprocal \gls{ris}, i.e., a solution to the right-hand side of \eqref{eq:perf}.
Since $f$ depends on $\mathbf{\Theta}$ only through the matrix $\mathbf{\Theta}\mathbf{H}_T$, if a reciprocal \gls{ris} $\mathbf{\Theta}_S^\star\in\mathcal{S}$ can produce the same matrix, i.e., $\mathbf{\Theta}_S^\star\mathbf{H}_T=\mathbf{\Theta}_U^\star\mathbf{H}_T$, then it achieves optimal performance, and \eqref{eq:perf} holds.
Therefore, comparing reciprocal and non-reciprocal \gls{ris} reduces to answering the following question: which matrices $\mathbf{\Theta}\mathbf{H}_T$ can be produced with $\mathbf{\Theta}\in\mathcal{U}$, and which with $\mathbf{\Theta}\in\mathcal{S}$?
We answer this question for non-reciprocal \gls{ris} in Lemma~\ref{lem:u} and for reciprocal \gls{ris} in Lemma~\ref{lem:s}.
\begin{lemma}
\label{lem:u}
Let $\mathbf{X},\mathbf{Y}\in\mathbb{C}^{M\times K}$.
There exists $\mathbf{\Theta}\in\mathcal{U}$ such that $\mathbf{\Theta}\mathbf{X}=\mathbf{Y}$ if and only if
\begin{equation}
\mathbf{X}^H\mathbf{X}=\mathbf{Y}^H\mathbf{Y}.\label{eq:u}
\end{equation}
\end{lemma}
\begin{proof}
Although Lemma~\ref{lem:u} is a classical result \cite[Theorem~7.3.11]{hor12}, we report its proof for completeness, as its structure is mirrored in the proof of Lemma~\ref{lem:s}.

\textit{Necessity.}
If $\mathbf{\Theta}\mathbf{X}=\mathbf{Y}$ with $\mathbf{\Theta}^H\mathbf{\Theta}=\mathbf{I}_M$, then $\mathbf{Y}^H\mathbf{Y}=\mathbf{X}^H\mathbf{\Theta}^H\mathbf{\Theta}\mathbf{X}=\mathbf{X}^H\mathbf{X}$.

\textit{Sufficiency.}
We show that if \eqref{eq:u} holds, then it is always possible to construct a unitary $\mathbf{\Theta}$ such that $\mathbf{\Theta}\mathbf{X}=\mathbf{Y}$.
Let $r$ be the rank of $\mathbf{X}$ and write its \gls{svd} as $\mathbf{X}=\mathbf{U}_1\mathbf{\Sigma}\mathbf{V}_1^H$, where $\mathbf{U}_1\in\mathbb{C}^{M\times r}$ and $\mathbf{V}_1\in\mathbb{C}^{K\times r}$ have orthonormal columns and $\mathbf{\Sigma}\in\mathbb{R}^{r\times r}$ is diagonal with positive entries.
Complete $\mathbf{U}_1$ to an $M\times M$ unitary matrix $\mathbf{U}=[\mathbf{U}_1,\mathbf{U}_2]$.

\textit{Step 1 (introduce a unitary $\mathbf{W}$).}
Define $\mathbf{W}_1=\mathbf{Y}\mathbf{V}_1\mathbf{\Sigma}^{-1}\in\mathbb{C}^{M\times r}$.
We have
\begin{align}
\mathbf{W}_1^H\mathbf{W}_1
&=\mathbf{\Sigma}^{-1}\mathbf{V}_1^H\mathbf{Y}^H\mathbf{Y}\mathbf{V}_1\mathbf{\Sigma}^{-1}\\
&\overset{(a)}{=}\mathbf{\Sigma}^{-1}\mathbf{V}_1^H\mathbf{X}^H\mathbf{X}\mathbf{V}_1\mathbf{\Sigma}^{-1}\\
&\overset{(b)}{=}\mathbf{\Sigma}^{-1}\mathbf{V}_1^H\mathbf{V}_1\mathbf{\Sigma}^2\mathbf{V}_1^H\mathbf{V}_1\mathbf{\Sigma}^{-1}\\
&\overset{(c)}{=}\mathbf{I}_r,
\end{align}
where $\overset{(a)}{=}$ follows from $\mathbf{Y}^H\mathbf{Y}=\mathbf{X}^H\mathbf{X}$, $\overset{(b)}{=}$ follows from $\mathbf{X}^H\mathbf{X}=\mathbf{V}_1\mathbf{\Sigma}^2\mathbf{V}_1^H$, and $\overset{(c)}{=}$ follows from $\mathbf{V}_1^H\mathbf{V}_1=\mathbf{I}_r$.
Therefore, $\mathbf{W}_1$ has orthonormal columns, and we can complete it to an $M\times M$ unitary matrix $\mathbf{W}=[\mathbf{W}_1,\mathbf{W}_2]$.

\textit{Step 2 (construct a unitary $\mathbf{\Theta}$).}
Define
\begin{equation}
\mathbf{\Theta}=\mathbf{W}\mathbf{U}^H=\mathbf{W}_1\mathbf{U}_1^H+\mathbf{W}_2\mathbf{U}_2^H,
\end{equation}
which is unitary as a product of unitary matrices.

\textit{Step 3 (verify that $\mathbf{\Theta}\mathbf{X}=\mathbf{Y}$).}
Since $\mathbf{U}_1^H\mathbf{U}_1=\mathbf{I}_r$ and $\mathbf{U}_2^H\mathbf{U}_1=\mathbf{0}$, we have
\begin{equation}
\mathbf{\Theta}\mathbf{X}=\mathbf{W}_1\mathbf{\Sigma}\mathbf{V}_1^H=\mathbf{Y}\mathbf{V}_1\mathbf{V}_1^H.
\end{equation}
It remains to show $\mathbf{Y}\mathbf{V}_1\mathbf{V}_1^H=\mathbf{Y}$.
Let $\mathbf{P}=\mathbf{I}_K-\mathbf{V}_1\mathbf{V}_1^H$, so that $\mathbf{X}\mathbf{P}=\mathbf{0}$.
Using $\mathbf{Y}^H\mathbf{Y}=\mathbf{X}^H\mathbf{X}$, we obtain
\begin{align}
\left(\mathbf{Y}\mathbf{P}\right)^H\left(\mathbf{Y}\mathbf{P}\right)
&=\mathbf{P}\mathbf{Y}^H\mathbf{Y}\mathbf{P}\\
&=\mathbf{P}\mathbf{X}^H\mathbf{X}\mathbf{P}\\
&=\left(\mathbf{X}\mathbf{P}\right)^H\left(\mathbf{X}\mathbf{P}\right)
=\mathbf{0}.
\end{align}
Hence $\mathbf{Y}\mathbf{P}=\mathbf{0}$, which gives $\mathbf{Y}\mathbf{V}_1\mathbf{V}_1^H=\mathbf{Y}$.
\end{proof}

Lemma~\ref{lem:u} states that a unitary $\mathbf{\Theta}$ can map the columns of $\mathbf{X}=[\mathbf{x}_1,\ldots,\mathbf{x}_K]$ to the columns of $\mathbf{Y}=[\mathbf{y}_1,\ldots,\mathbf{y}_K]$ if and only if all their norms and inner products are preserved, i.e., $\mathbf{x}_{k_1}^H\mathbf{x}_{k_2}=\mathbf{y}_{k_1}^H\mathbf{y}_{k_2}$, for $k_1,k_2=1,\ldots,K$.
For $K=1$, condition \eqref{eq:u} reduces to $\Vert\mathbf{x}\Vert=\Vert\mathbf{y}\Vert$.
Hence, a unitary $\mathbf{\Theta}$ can map any vector $\mathbf{x}$ to any vector $\mathbf{y}$ with the same norm.
For $K\geq2$, it is instructive to consider the case where the columns of $\mathbf{X}$ are orthogonal, so that $\mathbf{X}^H\mathbf{X}$ is diagonal.
In this case, \eqref{eq:u} holds if and only if the columns of $\mathbf{Y}$ are also orthogonal and $\Vert\mathbf{x}_k\Vert=\Vert\mathbf{y}_k\Vert$ for $k=1,\ldots,K$.
Hence, a unitary $\mathbf{\Theta}$ can map an arbitrary set of orthogonal vectors to an arbitrary set of orthogonal vectors with the same norms.
We now show how this flexibility is reduced when $\mathbf{\Theta}$ is also required to be symmetric.

\begin{lemma}
Let $\mathbf{X},\mathbf{Y}\in\mathbb{C}^{M\times K}$.
There exists $\mathbf{\Theta}\in\mathcal{S}$ such that $\mathbf{\Theta}\mathbf{X}=\mathbf{Y}$ if and only if
\begin{subequations}
\begin{align}
\mathbf{X}^H\mathbf{X}&=\mathbf{Y}^H\mathbf{Y},\label{eq:sa}\\
\mathbf{X}^T\mathbf{Y}&=\left(\mathbf{X}^T\mathbf{Y}\right)^T.\label{eq:sb}
\end{align}
\label{eq:s}
\end{subequations}
\label{lem:s}
\end{lemma}
\begin{proof}
This lemma implicitly appears in the proof of \cite[Theorem~2]{wu25} (see \cite[Section~IV-B]{wu25}).
Here, we report a self-contained proof that constructs a $\mathbf{\Theta}\in\mathcal{S}$ solving $\mathbf{\Theta}\mathbf{X}=\mathbf{Y}$, similar to the proof of Lemma~\ref{lem:u}.

\textit{Necessity.}
If $\mathbf{\Theta}\mathbf{X}=\mathbf{Y}$ with $\mathbf{\Theta}^H\mathbf{\Theta}=\mathbf{I}_M$, then $\mathbf{Y}^H\mathbf{Y}=\mathbf{X}^H\mathbf{\Theta}^H\mathbf{\Theta}\mathbf{X}=\mathbf{X}^H\mathbf{X}$.
If, in addition, $\mathbf{\Theta}=\mathbf{\Theta}^T$, then $\mathbf{X}^T\mathbf{Y}=\mathbf{X}^T\mathbf{\Theta}\mathbf{X}$ is symmetric.

\textit{Sufficiency.}
We show that if \eqref{eq:s} holds, then it is always possible to construct a symmetric unitary $\mathbf{\Theta}$ such that $\mathbf{\Theta}\mathbf{X}=\mathbf{Y}$.
To this end, we exploit the fact that $\mathbf{\Theta}\mathbf{X}=\mathbf{Y}$ also implies $\mathbf{\Theta}\mathbf{Y}^*=\mathbf{X}^*$ when $\mathbf{\Theta}$ is symmetric unitary, since $\mathbf{\Theta}^{-1}=\mathbf{\Theta}^*$.
We therefore construct a symmetric unitary $\mathbf{\Theta}$ satisfying both equations at once, i.e., $\mathbf{\Theta}\tilde{\mathbf{X}}=\tilde{\mathbf{Y}}$, where $\tilde{\mathbf{X}}=\left[\mathbf{X},\mathbf{Y}^*\right]$, $\tilde{\mathbf{Y}}=\left[\mathbf{Y},\mathbf{X}^*\right]\in\mathbb{C}^{M\times 2K}$.
Let $r$ be the rank of $\tilde{\mathbf{X}}$ and write its \gls{svd} as $\tilde{\mathbf{X}}=\mathbf{U}_1\mathbf{\Sigma}\mathbf{V}_1^H$, where $\mathbf{U}_1\in\mathbb{C}^{M\times r}$ and $\mathbf{V}_1\in\mathbb{C}^{2K\times r}$ have orthonormal columns and $\mathbf{\Sigma}\in\mathbb{R}^{r\times r}$ is diagonal with positive entries.
Complete $\mathbf{U}_1$ to an $M\times M$ unitary matrix $\mathbf{U}=[\mathbf{U}_1,\mathbf{U}_2]$.
Introducing the symmetric permutation matrix
\begin{equation}
\mathbf{\Pi}=
\begin{bmatrix}
\mathbf{0} & \mathbf{I}_K\\
\mathbf{I}_K & \mathbf{0}
\end{bmatrix},
\end{equation}
 we have
\begin{equation}
\tilde{\mathbf{Y}}=\tilde{\mathbf{X}}^*\mathbf{\Pi}=\mathbf{U}_1^*\mathbf{\Sigma}\mathbf{V}_1^T\mathbf{\Pi}.
\end{equation}
From the definitions of $\tilde{\mathbf{X}}$ and $\tilde{\mathbf{Y}}$, if \eqref{eq:s} holds, we have
\begin{equation}
\tilde{\mathbf{X}}^H\tilde{\mathbf{X}}=\tilde{\mathbf{Y}}^H\tilde{\mathbf{Y}}.\label{eq:gram}
\end{equation}
Substituting the \gls{svd} expressions of $\tilde{\mathbf{X}}$ and $\tilde{\mathbf{Y}}$ into \eqref{eq:gram} gives
\begin{equation}
\mathbf{V}_1\mathbf{\Sigma}^2\mathbf{V}_1^H=\mathbf{\Pi}\mathbf{V}_1^*\mathbf{\Sigma}^2\mathbf{V}_1^T\mathbf{\Pi}.\label{eq:gram2}
\end{equation}

\textit{Step 1 (introduce a symmetric unitary $\mathbf{C}$).}
Define the $r\times r$ matrices
\begin{equation}
\mathbf{N}=\mathbf{V}_1^T\mathbf{\Pi}\mathbf{V}_1,\;\;\;
\mathbf{C}=\mathbf{\Sigma}\mathbf{N}\mathbf{\Sigma}^{-1}.
\end{equation}
$\mathbf{N}$ is symmetric because $\mathbf{\Pi}$ is symmetric.
In the following, we show that $\mathbf{C}$ is symmetric unitary.
Multiplying \eqref{eq:gram2} by $\mathbf{V}_1^T\mathbf{\Pi}$ on the left and by $\mathbf{V}_1$ on the right gives $\mathbf{N}\mathbf{\Sigma}^2=\mathbf{\Sigma}^2\mathbf{N}$.
Multiplying it on the left by $\mathbf{\Sigma}^{-1}$ and on the right by $\mathbf{\Sigma}^{-1}$ yields $\mathbf{\Sigma}^{-1}\mathbf{N}\mathbf{\Sigma}=\mathbf{\Sigma}\mathbf{N}\mathbf{\Sigma}^{-1}$, showing that $\mathbf{C}$ is symmetric.
Furthermore, multiplying \eqref{eq:gram2} by $\mathbf{V}_1^H$ on the left and by $\mathbf{V}_1$ on the right gives $\mathbf{\Sigma}^2=\mathbf{N}^H\mathbf{\Sigma}^2\mathbf{N}$.
Thus, $\mathbf{C}^H\mathbf{C}=\mathbf{\Sigma}^{-1}\mathbf{N}^H\mathbf{\Sigma}^2\mathbf{N}\mathbf{\Sigma}^{-1}=\mathbf{I}_r$, proving that $\mathbf{C}$ is unitary.

\textit{Step 2 (construct a symmetric unitary $\mathbf{\Theta}$).}
Define
\begin{equation}
\mathbf{\Theta}=\mathbf{U}^*
\begin{bmatrix}
\mathbf{C} & \mathbf{0}\\
\mathbf{0} & \mathbf{I}_{M-r}
\end{bmatrix}
\mathbf{U}^H,
\end{equation}
which is unitary as a product of unitary matrices, and symmetric because the middle matrix is symmetric and $\mathbf{U}^H=(\mathbf{U}^*)^T$.

\textit{Step 3 (verify that $\mathbf{\Theta}\mathbf{X}=\mathbf{Y}$).}
Since $\mathbf{U}^H\mathbf{U}_1=[\mathbf{I}_r,\mathbf{0}]^T$, we have
\begin{equation}
\mathbf{\Theta}\tilde{\mathbf{X}}
=\mathbf{U}_1^*\mathbf{C}\mathbf{\Sigma}\mathbf{V}_1^H
=\mathbf{U}_1^*\mathbf{\Sigma}\mathbf{V}_1^T\mathbf{\Pi}\mathbf{V}_1\mathbf{V}_1^H
=\tilde{\mathbf{Y}}\mathbf{V}_1\mathbf{V}_1^H.
\end{equation}
It remains to show $\tilde{\mathbf{Y}}\mathbf{V}_1\mathbf{V}_1^H=\tilde{\mathbf{Y}}$.
Let $\mathbf{P}=\mathbf{I}_{2K}-\mathbf{V}_1\mathbf{V}_1^H$, so that $\tilde{\mathbf{X}}\mathbf{P}=\mathbf{0}$.
Using \eqref{eq:gram}, we get $(\tilde{\mathbf{Y}}\mathbf{P})^H(\tilde{\mathbf{Y}}\mathbf{P})=\mathbf{0}$.
Hence, $\tilde{\mathbf{Y}}\mathbf{P}=\mathbf{0}$, which gives $\tilde{\mathbf{Y}}\mathbf{V}_1\mathbf{V}_1^H=\tilde{\mathbf{Y}}$.
We have finally obtained $\mathbf{\Theta}\tilde{\mathbf{X}}=\tilde{\mathbf{Y}}$, whose first $K$ columns are $\mathbf{\Theta}\mathbf{X}=\mathbf{Y}$.
\end{proof}

Lemma~\ref{lem:s} states that $\mathbf{\Theta}\mathbf{X}=\mathbf{Y}$ has a symmetric unitary solution if and only if it satisfies, in addition to \eqref{eq:sa}, also the symmetry condition \eqref{eq:sb}, i.e., $\mathbf{x}_{k_1}^T\mathbf{y}_{k_2}=\mathbf{x}_{k_2}^T\mathbf{y}_{k_1}$, for $k_1,k_2=1,\ldots,K$.
For $K=1$, condition \eqref{eq:sa} reduces to $\Vert\mathbf{x}\Vert=\Vert\mathbf{y}\Vert$, and \eqref{eq:sb} is a scalar identity that always holds.
Hence, a symmetric unitary $\mathbf{\Theta}$ can map any vector $\mathbf{x}$ to any vector $\mathbf{y}$ with the same norm, exactly as a unitary $\mathbf{\Theta}$.
For $K\geq2$, \eqref{eq:sb} imposes $K(K-1)/2$ additional constraints.
Therefore, unlike a unitary $\mathbf{\Theta}$, a symmetric unitary $\mathbf{\Theta}$ cannot map an arbitrary set of orthogonal vectors to an arbitrary set of orthogonal vectors with the same norms.

\section{A Necessary and Sufficient Condition}

In this section, we derive a necessary and sufficient condition under which reciprocal \gls{ris} performs as non-reciprocal \gls{ris}.

Notice that reciprocal \gls{ris} performs as non-reciprocal \gls{ris} when it can reproduce the matrix $\mathbf{\Theta}^\star\mathbf{H}_T$ produced by an optimal non-reciprocal \gls{ris} $\mathbf{\Theta}^\star$.
Lemma~\ref{lem:s} states when this is possible.
Introducing $\mathbf{X}=\mathbf{H}_T$ and $\mathbf{Y}=\mathbf{\Theta}^\star\mathbf{H}_T$, condition \eqref{eq:sa} is always satisfied since $\mathbf{\Theta}^\star$ is unitary.
Therefore, the only obstacle for a reciprocal \gls{ris} is condition \eqref{eq:sb}, which requires $\mathbf{X}^T\mathbf{Y}=\mathbf{H}_T^T\mathbf{\Theta}^\star\mathbf{H}_T$ to be symmetric.
This observation is formalized in the following theorem.

\begin{theorem}
\label{thm:main}
Reciprocal \gls{ris} performs as non-reciprocal \gls{ris} if and only if there exists an optimal non-reciprocal \gls{ris} $\mathbf{\Theta}^\star\in\mathcal{U}$ for which $\mathbf{H}_T^T\mathbf{\Theta}^\star\mathbf{H}_T$ is symmetric.
\end{theorem}
\begin{proof}
\textit{Necessity.}
If \eqref{eq:perf} holds, $\mathbf{\Theta}^\star\in\mathcal{S}$ is an optimal non-reciprocal \gls{ris} (since $\mathcal{S}\subset\mathcal{U}$), and $\mathbf{H}_T^T\mathbf{\Theta}^\star\mathbf{H}_T$ is symmetric because $\mathbf{\Theta}^\star$ is.

\textit{Sufficiency.}
Assume there exists an optimal non-reciprocal \gls{ris} $\mathbf{\Theta}_U^\star\in\mathcal{U}$ such that $\mathbf{H}_T^T\mathbf{\Theta}_U^\star\mathbf{H}_T$ is symmetric.
We show that a reciprocal \gls{ris} can produce the same matrix $\mathbf{\Theta}\mathbf{H}_T$, and therefore achieve optimal performance.
To this end, consider $\mathbf{X}=\mathbf{H}_T$ and $\mathbf{Y}=\mathbf{\Theta}_U^\star\mathbf{H}_T$.
Since $\mathbf{\Theta}_U^\star$ is unitary, we have $\mathbf{Y}^H\mathbf{Y}=\mathbf{H}_T^H\mathbf{H}_T=\mathbf{X}^H\mathbf{X}$.
In addition, $\mathbf{X}^T\mathbf{Y}=\mathbf{H}_T^T\mathbf{\Theta}_U^\star\mathbf{H}_T$ is symmetric.
Therefore, by Lemma~\ref{lem:s}, there exists a reciprocal \gls{ris} $\mathbf{\Theta}_S^\star\in\mathcal{S}$ such that $\mathbf{\Theta}_S^\star\mathbf{H}_T=\mathbf{\Theta}_U^\star\mathbf{H}_T$, achieving optimal performance.
\end{proof}

Theorem~\ref{thm:main} can be physically interpreted as follows.
Since $\mathbf{H}_T^T$ is the channel from the \gls{ris} back to the transmitting antennas, $\mathbf{H}_T^T\mathbf{\Theta}\mathbf{H}_T$ is the round-trip channel from the transmitting antennas, via the \gls{ris}, back to the transmitting antennas, as shown in Fig.~\ref{fig:system}(b).
Its $(k_1,k_2)$th entry is the channel from the $k_2$th to the $k_1$th transmitting antenna via the \gls{ris}, so its symmetry means that this channel is the same in both directions, i.e., the round-trip channel is reciprocal.
A reciprocal \gls{ris} can never break this reciprocity, and Theorem~\ref{thm:main} shows that this is the only limitation of reciprocal \gls{ris} that matters.
In other words, reciprocal \gls{ris} performs as non-reciprocal \gls{ris} if and only if optimal performance does not require breaking the reciprocity of this round-trip channel.
Note that for metrics of the form $f(\mathbf{H}_R\mathbf{\Theta})$, an analogous result holds with $\mathbf{H}_R\mathbf{\Theta}^\star\mathbf{H}_R^T$ in place of $\mathbf{H}_T^T\mathbf{\Theta}^\star\mathbf{H}_T$, i.e., the round-trip channel from the receiving antennas, via the \gls{ris}, back to the receiving antennas.
For metrics that are functions of $\mathbf{H}_R\mathbf{\Theta}\mathbf{H}_T$, which are of both forms $f(\mathbf{\Theta}\mathbf{H}_T)$ and $f(\mathbf{H}_R\mathbf{\Theta})$, either of the two conditions can be equivalently used.

A notable case where the round-trip channel $\mathbf{H}_T^T\mathbf{\Theta}\mathbf{H}_T$ is always symmetric is when $\mathbf{H}_T$ has rank one, as stated in the following corollary.
\begin{corollary}
\label{cor:rank1}
If $\mathrm{rank}(\mathbf{H}_T)=1$, then reciprocal \gls{ris} performs as non-reciprocal \gls{ris}.
\end{corollary}
\begin{proof}
If $\mathrm{rank}(\mathbf{H}_T)=1$, we have $\mathbf{H}_T=\mathbf{a}\mathbf{b}^T$ for some
$\mathbf{a}\in\mathbb{C}^{M\times 1}$ and $\mathbf{b}\in\mathbb{C}^{N_T\times 1}$,
so that
\begin{equation}
\mathbf{H}_T^T\mathbf{\Theta}\mathbf{H}_T
=\mathbf{b}\left(\mathbf{a}^T\mathbf{\Theta}\mathbf{a}\right)\mathbf{b}^T
\overset{(a)}{=}\left(\mathbf{a}^T\mathbf{\Theta}\mathbf{a}\right)\mathbf{b}\mathbf{b}^T,
\end{equation}
where $\overset{(a)}{=}$ follows from $\mathbf{a}^T\mathbf{\Theta}\mathbf{a}$ being a scalar.
Since $\mathbf{H}_T^T\mathbf{\Theta}\mathbf{H}_T$ is symmetric, this corollary follows from Theorem~\ref{thm:main}.
\end{proof}

Corollary~\ref{cor:rank1} includes all systems with a single transmitting antenna ($N_T=1$), regardless of the number of receiving antennas.
It also includes all systems where all transmitting antennas belong to one device in \gls{los} and far-field conditions with the \gls{ris}, since $\mathbf{H}_T$ is then the outer product of the steering vectors of the \gls{ris} and of the device.
Similarly, for metrics of the form $f(\mathbf{H}_R\mathbf{\Theta})$, the same holds if $\mathrm{rank}(\mathbf{H}_R)=1$, e.g., with a single receiving antenna ($N_R=1$) or a single receiving device in \gls{los} and far-field conditions with the \gls{ris}.
For metrics that are functions of $\mathbf{H}_R\mathbf{\Theta}\mathbf{H}_T$, reciprocal \gls{ris} performs as non-reciprocal \gls{ris} if either $\mathbf{H}_T$ or $\mathbf{H}_R$ has rank one.
Hence, single-user \gls{siso}, \gls{simo}, and \gls{miso} systems never benefit from non-reciprocal \gls{ris}.

\section{A Necessary and Sufficient Condition\\in the Presence of Precoding}

In this section, we refine the condition in Theorem~\ref{thm:main} by accounting for the transmitter precoder.

Consider a transmitted signal $\mathbf{x}=\mathbf{F}\mathbf{s}$, where $\mathbf{s}\in\mathbb{C}^{K\times 1}$ contains $K\leq N_T$ symbols (data streams) and $\mathbf{F}\in\mathcal{F}\subseteq\mathbb{C}^{N_T\times K}$ is the precoder, with $\mathcal{F}$ being the set of feasible precoders, e.g., satisfying a power constraint.
Since the received signal is $\mathbf{y}=\mathbf{H}\mathbf{F}\mathbf{s}+\mathbf{n}$, we consider metrics that depend on $\mathbf{\Theta}$ through $\mathbf{\Theta}\mathbf{H}_T\mathbf{F}$.
We therefore write them as $g(\mathbf{F},\mathbf{\Theta}\mathbf{H}_T\mathbf{F})$, where $g$ is real-valued and can also depend on $\mathbf{F}$ directly, e.g., through $\mathbf{H}_{RT}\mathbf{F}$.
We assume that the maximum of the metric over $\mathcal{U}\times\mathcal{F}$ is attained, which holds, e.g., when $g$ is continuous and $\mathcal{F}$ is compact.
\begin{definition}
\label{def:perf-F}
We say that reciprocal \gls{ris} performs as non-reciprocal \gls{ris} in the presence of precoding when there exists a pair $(\mathbf{\Theta}^\star,\mathbf{F}^\star)\in\mathcal{S}\times\mathcal{F}$ such that
\begin{equation}
g\left(\mathbf{F}^\star,\mathbf{\Theta}^\star\mathbf{H}_T\mathbf{F}^\star\right)=
\max_{\mathbf{\Theta}\in\mathcal{U},\mathbf{F}\in\mathcal{F}} g\left(\mathbf{F},\mathbf{\Theta}\mathbf{H}_T\mathbf{F}\right).\label{eq:perf-F}
\end{equation}
\end{definition}

Note that a reciprocal \gls{ris} now only needs to reproduce the matrix $\mathbf{\Theta}^\star\mathbf{H}_T\mathbf{F}^\star$ produced by an optimal pair of non-reciprocal \gls{ris} $\mathbf{\Theta}^\star$ and precoder $\mathbf{F}^\star$, rather than the whole matrix $\mathbf{\Theta}^\star\mathbf{H}_T$.
This leads to the following necessary and sufficient condition, which is weaker than the one in Theorem~\ref{thm:main}.
\begin{theorem}
\label{thm:precoder}
Reciprocal \gls{ris} performs as non-reciprocal \gls{ris} in the presence of precoding if and only if there exists an optimal pair of non-reciprocal \gls{ris} $\mathbf{\Theta}^\star\in\mathcal{U}$ and precoder $\mathbf{F}^\star\in\mathcal{F}$ for which $\mathbf{F}^{\star T}\mathbf{H}_T^T\mathbf{\Theta}^\star\mathbf{H}_T\mathbf{F}^\star$ is symmetric.
\end{theorem}
\begin{proof}
\textit{Necessity.}
If \eqref{eq:perf-F} holds, $(\mathbf{\Theta}^\star,\mathbf{F}^\star)\in\mathcal{S}\times\mathcal{F}$ is an optimal pair for the non-reciprocal \gls{ris} problem (since $\mathcal{S}\subset\mathcal{U}$), and $\mathbf{F}^{\star T}\mathbf{H}_T^T\mathbf{\Theta}^\star\mathbf{H}_T\mathbf{F}^\star$ is symmetric because $\mathbf{\Theta}^\star$ is.

\textit{Sufficiency.}
Assume there exists an optimal pair $(\mathbf{\Theta}_U^\star,\mathbf{F}^\star)\in\mathcal{U}\times\mathcal{F}$ of the non-reciprocal \gls{ris} problem such that $\mathbf{F}^{\star T}\mathbf{H}_T^T\mathbf{\Theta}_U^\star\mathbf{H}_T\mathbf{F}^\star$ is symmetric.
We show that a reciprocal \gls{ris} can produce the same matrix $\mathbf{\Theta}\mathbf{H}_T\mathbf{F}^\star$, and therefore achieve optimal performance with the same precoder.
To this end, consider $\mathbf{X}=\mathbf{H}_T\mathbf{F}^\star$ and $\mathbf{Y}=\mathbf{\Theta}_U^\star\mathbf{H}_T\mathbf{F}^\star$.
Since $\mathbf{\Theta}_U^\star$ is unitary, we have $\mathbf{Y}^H\mathbf{Y}=\mathbf{X}^H\mathbf{X}$.
In addition, $\mathbf{X}^T\mathbf{Y}=\mathbf{F}^{\star T}\mathbf{H}_T^T\mathbf{\Theta}_U^\star\mathbf{H}_T\mathbf{F}^\star$ is symmetric.
Therefore, by Lemma~\ref{lem:s}, there exists a reciprocal \gls{ris} $\mathbf{\Theta}_S^\star\in\mathcal{S}$ such that $\mathbf{\Theta}_S^\star\mathbf{H}_T\mathbf{F}^\star=\mathbf{\Theta}_U^\star\mathbf{H}_T\mathbf{F}^\star$.
The pair $(\mathbf{\Theta}_S^\star,\mathbf{F}^\star)$ is feasible for the reciprocal \gls{ris} problem and achieves optimal performance.
\end{proof}

The condition in Theorem~\ref{thm:precoder} involves the $K\times K$ round-trip channel seen by the transmitted streams, i.e., from the $K$ streams, via the \gls{ris}, back to the $K$ streams, as shown in Fig.~\ref{fig:system}(c).
This is weaker than the condition in Theorem~\ref{thm:main}, since $\mathbf{F}^{\star T}\mathbf{M}\mathbf{F}^\star$ is symmetric whenever $\mathbf{M}$ is symmetric, but not vice versa.
Thus, a reciprocal \gls{ris} only needs to preserve the reciprocity of the round-trip channel along the directions actually used for transmission.
Similarly, if the receiving antennas use a combiner $\mathbf{G}^\star\in\mathbb{C}^{K\times N_R}$, Theorem~\ref{thm:precoder} holds with $\mathbf{G}^\star\mathbf{H}_R\mathbf{\Theta}^\star\mathbf{H}_R^T\mathbf{G}^{\star T}$ in place of
$\mathbf{F}^{\star T}\mathbf{H}_T^T\mathbf{\Theta}^\star\mathbf{H}_T\mathbf{F}^\star$.

A notable case where the condition of Theorem~\ref{thm:precoder} always holds is single-stream transmission, as stated in the following.
\begin{corollary}
\label{cor:single}
If single-stream transmission ($K=1$) is optimal with non-reciprocal \gls{ris}, then reciprocal \gls{ris} performs as non-reciprocal \gls{ris}.
\end{corollary}
\begin{proof}
Since $\mathbf{F}^{\star T}\mathbf{H}_T^T\mathbf{\Theta}^\star\mathbf{H}_T\mathbf{F}^\star$ is a scalar for $K=1$, and therefore symmetric, this corollary follows from Theorem~\ref{thm:precoder}.
\end{proof}

Corollary~\ref{cor:single} extends Corollary~\ref{cor:rank1} to systems where $\mathbf{H}_T$ and $\mathbf{H}_R$ have arbitrary rank but only one stream is transmitted.
This includes, for example, single-user \gls{mimo} systems at low \gls{snr} and multi-user \gls{miso} systems maximizing the sum-power, where transmission over the dominant channel eigenmode is optimal.
In addition, this also includes multi-user \gls{miso} systems maximizing the sum-rate at low \gls{snr}, where only the strongest user is served.
In these systems, reciprocal \gls{ris} performs as non-reciprocal \gls{ris}, consistent with the optimal reciprocal \gls{ris} solutions derived for single-stream systems in \cite{ner24,san23}.

\section{Conclusion}

We have answered the question of whether reciprocal \gls{ris} can perform as non-reciprocal \gls{ris}.
This happens if and only if there exists an optimal non-reciprocal \gls{ris} that makes the round-trip channel from the transmitting antennas, via the \gls{ris}, back to the transmitting antennas symmetric.
When the precoder is taken into account, this condition only needs to hold for the round-trip channel seen by the transmitted streams.
As a result, reciprocal \gls{ris} performs as non-reciprocal \gls{ris} when the channel from the transmitting antennas to the \gls{ris} has rank one or when single-stream transmission is optimal.
In these systems, the additional hardware complexity of non-reciprocal \gls{ris} brings no performance gain.

Conversely, non-reciprocal \gls{ris} is beneficial when optimal performance requires breaking the reciprocity of the round-trip channel.
The clearest example is a full-duplex system, where a base station simultaneously receives from an uplink user and transmits to a downlink user located in a different direction, both via the \gls{ris} \cite{li25,liu26-a,liu26-b,liu26-c}.
Here, a reciprocal \gls{ris} forces the channel from the uplink user to the base station to be equal to the reverse one, while optimal performance requires the former to be strong and the latter weak, since the base station signal should be reflected towards the downlink user.
More generally, non-reciprocal \gls{ris} is expected to be beneficial when numerous streams are transmitted through well-conditioned channels \cite{ale25}, e.g., when the \gls{ris} serves multiple transmitters and receivers in different directions, or in rich scattering environments.

\bibliographystyle{IEEEtran}
\bibliography{IEEEabrv,main}

\begin{thebibliography}{10}
\providecommand{\url}[1]{#1}
\csname url@samestyle\endcsname
\providecommand{\newblock}{\relax}
\providecommand{\bibinfo}[2]{#2}
\providecommand{\BIBentrySTDinterwordspacing}{\spaceskip=0pt\relax}
\providecommand{\BIBentryALTinterwordstretchfactor}{4}
\providecommand{\BIBentryALTinterwordspacing}{\spaceskip=\fontdimen2\font plus
\BIBentryALTinterwordstretchfactor\fontdimen3\font minus \fontdimen4\font\relax}
\providecommand{\BIBforeignlanguage}[2]{{%
\expandafter\ifx\csname l@#1\endcsname\relax
\typeout{** WARNING: IEEEtran.bst: No hyphenation pattern has been}%
\typeout{** loaded for the language `#1'. Using the pattern for}%
\typeout{** the default language instead.}%
\else
\language=\csname l@#1\endcsname
\fi
#2}}
\providecommand{\BIBdecl}{\relax}
\BIBdecl

\bibitem{wu21}
Q.~Wu, S.~Zhang, B.~Zheng, C.~You, and R.~Zhang, ``Intelligent reflecting surface-aided wireless communications: A tutorial,'' \emph{IEEE Trans. Commun.}, vol.~69, no.~5, pp. 3313--3351, 2021.

\bibitem{zha20}
S.~Zhang and R.~Zhang, ``Capacity characterization for intelligent reflecting surface aided {MIMO} communication,'' \emph{IEEE J. Sel. Areas Commun.}, vol.~38, no.~8, pp. 1823--1838, 2020.

\bibitem{poz11}
D.~M. Pozar, \emph{Microwave Engineering}.\hskip 1em plus 0.5em minus 0.4em\relax John wiley \& sons, 2011.

\bibitem{li26}
H.~Li, M.~Nerini, S.~Shen, and B.~Clerckx, ``A tutorial on beyond-diagonal reconfigurable intelligent surfaces: Modeling, architectures, system design and optimization, and applications,'' \emph{IEEE Commun. Surv. Tutor.}, vol.~28, pp. 4086--4126, 2026.

\bibitem{she22}
S.~Shen, B.~Clerckx, and R.~Murch, ``Modeling and architecture design of reconfigurable intelligent surfaces using scattering parameter network analysis,'' \emph{IEEE Trans. Wireless Commun.}, vol.~21, no.~2, pp. 1229--1243, 2022.

\bibitem{ale25}
J.~V. Alegría, J.~Thunberg, and O.~Edfors, ``Channel orthogonalization with reconfigurable surfaces: General models, theoretical limits, and effective configuration,'' \emph{IEEE Trans. Wireless Commun.}, vol.~24, no.~6, pp. 5179--5195, 2025.

\bibitem{bar23}
G.~Bartoli, A.~Abrardo, N.~Decarli, D.~Dardari, and M.~Di~Renzo, ``Spatial multiplexing in near field {MIMO} channels with reconfigurable intelligent surfaces,'' \emph{IET Signal Processing}, vol.~17, no.~3, p. e12195, 2023.

\bibitem{xu25}
J.~Xu, H.~Wang, R.~Liu, J.~A. Nossek, and A.~L. Swindlehurst, ``Non-reciprocal reconfigurable intelligent surfaces,'' \emph{IEEE Wireless Commun. Lett.}, vol.~14, no.~10, pp. 3329--3333, 2025.

\bibitem{wan24}
H.~Wang, Z.~Han, and A.~L. Swindlehurst, ``Channel reciprocity attacks using intelligent surfaces with non-diagonal phase shifts,'' \emph{IEEE Open J. Commun. Soc.}, vol.~5, pp. 1469--1485, 2024.

\bibitem{wan26}
H.~Wang, J.~Hu, J.~Xu, Y.~Ju, and A.~L. Swindlehurst, ``{DRL}-based mitigation against nonreciprocal {RIS}-aided channel reciprocity attacks,'' in \emph{ICC 2026 - IEEE International Conference on Communications}, 2026.

\bibitem{wan25}
------, ``Nonreciprocal {RIS}-aided covert channel reciprocity attacks and countermeasures,'' \emph{arXiv preprint arXiv:2509.11117}, 2025.

\bibitem{li25}
H.~Li and B.~Clerckx, ``Non-reciprocal beyond diagonal {RIS}: Multiport network models and performance benefits in full-duplex systems,'' \emph{IEEE Trans. Commun.}, vol.~73, no.~11, pp. 12\,221--12\,234, 2025.

\bibitem{liu26-a}
Z.~Liu, H.~Li, and B.~Clerckx, ``Non-reciprocal beyond diagonal {RIS}: Sum-rate maximization in full-duplex communications,'' \emph{IEEE Trans. Commun.}, vol.~74, pp. 5169--5183, 2026.

\bibitem{liu26-b}
Z.~Liu and B.~Clerckx, ``A secure full-duplex wireless circulator enabled by non-reciprocal beyond-diagonal {RIS},'' \emph{IEEE J. Sel. Areas Commun.}, vol.~44, pp. 3750--3765, 2026.

\bibitem{liu26-c}
Z.~Liu, W.~Shin, and B.~Clerckx, ``Enabling full-duplex {LEO} satellite systems with non-reciprocal {BD-RIS}-assisted beamforming,'' \emph{IEEE Trans. Wireless Commun.}, vol.~25, pp. 15\,459--15\,473, 2026.

\bibitem{wu26}
Z.~Wu, M.~Nerini, and B.~Clerckx, ``{RIS} optimization and scaling laws in multi-operator systems: Is quadratic scaling achievable?'' \emph{arXiv preprint arXiv:2605.01389}, 2026.

\bibitem{hor12}
R.~A. Horn and C.~R. Johnson, \emph{Matrix analysis}.\hskip 1em plus 0.5em minus 0.4em\relax Cambridge University Press, 2012.

\bibitem{wu25}
Z.~Wu and B.~Clerckx, ``Beyond-diagonal {RIS} in multiuser {MIMO}: Graph theoretic modeling and optimal architectures with low complexity,'' \emph{IEEE Trans. Inf. Theory}, vol.~71, no.~11, pp. 8506--8523, 2025.

\bibitem{ner24}
M.~Nerini, S.~Shen, and B.~Clerckx, ``Closed-form global optimization of beyond diagonal reconfigurable intelligent surfaces,'' \emph{IEEE Trans. Wireless Commun.}, vol.~23, no.~2, pp. 1037--1051, 2024.

\bibitem{san23}
I.~Santamaria, M.~Soleymani, E.~Jorswieck, and J.~Gutiérrez, ``{SNR} maximization in beyond diagonal {RIS}-assisted single and multiple antenna links,'' \emph{IEEE Signal Process. Lett.}, vol.~30, pp. 923--926, 2023.

\end{thebibliography}

\end{document}